\documentclass[12pt]{article}

\usepackage{amsmath}
\usepackage{amsfonts}
\usepackage{amssymb}
\usepackage{graphicx}
\usepackage{psfrag}
\usepackage{graphicx}
\usepackage{psfrag}

\newcommand{\be}{\begin{equation}}
\newcommand{\ee}{\end{equation}}
\newcommand{\ba}{\begin{eqnarray}}
\newcommand{\ea}{\end{eqnarray}}

\begin{document}
\begin{center}
{\bf
THREE-DIMENSIONAL SHAPE INVARIANT NON-SEPARABLE MODEL WITH EQUIDISTANT SPECTRUM}\\
\vspace{1cm}
{\large \bf M. S. Bardavelidze$^{1,}$\footnote{E-mail: m.bardaveli@gmail.com}, F. Cannata$^{2,}$\footnote{E-mail: cannata@bo.infn.it},
M. V. Iof\/fe$^{3,}$\footnote{E-mail: m.ioffe@pobox.spbu.ru},
D. N. Nishnianidze}$^{1,3,}$\footnote{E-mail: cutaisi@yahoo.com}\\
\vspace{0.5cm}
$^1$ Akaki Tsereteli State University, 4600 Kutaisi, Georgia\\
$^2$ INFN, Via Irnerio 46, 40126 Bologna, Italy.\\
$^3$ Saint-Petersburg State University,198504 Sankt-Petersburg, Russia
\end{center}
\vspace{0.5cm}
\hspace*{0.5in}
\vspace{1cm}
\hspace*{0.5in}
\begin{minipage}{5.0in}
{\small
A class of three-dimensional models which satisfy supersymmetric intertwining relations with the simplest - oscillator-like - variant of shape invariance is constructed. It is proved that the models are not amenable to conventional separation of variables for the complex potentials, but their spectra are real and equidistant (like for isotropic harmonic oscillator). The special case of such potential with quadratic interaction is solved completely. The Hamiltonian of the system is non-diagonalizable, and its wave functions and associated functions are built analytically. The symmetry properties of the model and degeneracy of energy levels are studied.}
\end{minipage}


\section*{\normalsize\bf 1. \quad Introduction.}

The idea of shape invariance was proposed \cite{gendenshtein} in the framework of one-dimensional Supersymmetrical Quantum Mechanics \cite{witten}, \cite{cooper} providing a very elegant method of complete solution of a class of one-dimensional quantum models. Although almost all these models were already solved, this approach gave a new (purely algebraic) look on the class of exactly solvable models. The Hamiltonian $H^{(1)}(x; a)$ depending on a space variable $x$ and a constant parameter $\gamma$ is said to be shape invariant iff: 1) it satisfies supersymmetric intertwining relations:
\begin{equation}\label{int}
H^{(1)}(x; \gamma ) q(x) = q(x) H^{(2)}(x; \gamma ),
\end{equation}
with some intertwining operators $q,$ and 2) two Hamiltonians in (\ref{int}) are related as follows:
\begin{equation}\label{shape}
H^{(2)}(x; \widetilde \gamma ) = H^{(1)}(x; \gamma ) + R(\gamma ),
\end{equation}
where $R(\gamma )$ does not depend on coordinate $x,$ and the parameters $\gamma $ and $\widetilde \gamma $ are
related: $\widetilde \gamma  = \widetilde \gamma (\gamma ).$ Eq.(\ref{shape}) just means that intertwined Hamiltonians have the same shape, and they differ only a little bit due to a change of parameter. The general form of shape invariance was investigated in \cite{mallow}, where all shape invariant potentials with additive dependence of $\widetilde\gamma$ on $\gamma$ were found.

The method of shape invariance was generalized also to the two-dimensional Schr\"odinger operators not amenable to conventional separation of variables in \cite{new}, \cite{ioffe1}. It was used in formulation of two new - supersymmetric - methods of separation of variables. One of them \cite{new}, \cite{ioffe1} was useful in the study of several quasi-exactly solvable models where a part of spectrum and corresponding wave functions was found analytically for arbitrary values of parameters of the models. The second one \cite{morse}, \cite{ioffe2} provided complete solution of these models, but only for specific values of parameters. The general form of two-dimensional systems with additive shape invariance was found in \cite{shape-2011}. In particular, the new shape invariant potential (two-dimensional generalization of Scarf II) was built, and its spectrum was constructed \cite{scarf} by the method of supersymmetric separation of variables mentioned above.

Of special interest are the models with the simplest kind of shape invariance with
\begin{equation}\label{osc-shape}
\widetilde \gamma  \equiv \gamma ; \quad R(\gamma )=2\lambda .
\end{equation}
Just this property with intertwining operators $q$ of first order in derivatives is satisfied by one-dimensional harmonic oscillator leading to the well known equidistant spectrum with the spacing $2\lambda .$ Much more interesting models with the same properties (\ref{osc-shape}) but with higher order intertwining operators were studied in \cite{shape2000}. It was shown that depending on the structure of zero modes of higher order intertwining operators $q$ and $q^{\dagger}$ the spectrum of the model consists of finite or infinite sequences of equidistant levels with the spacing $2\lambda .$ The systems with shape invariance of the form (\ref{osc-shape}) are naturally called "oscillator-like" shape invariant (sometimes also called "self-isospectral").

In two-dimensional space, the oscillator-like shape invariance (\ref{osc-shape}) with first order differential operators $q$ is fulfilled in a nontrivial model with quadratic interaction with such a choice of coupling constants that standard separation of variables is impossible (see \cite{osc}). The Hamiltonian of this model is non-Hermitian and non-diagonalizable, but its spectrum is the infinite equidistant sequence of real energy levels with the same spacing $2\lambda .$ Moreover, the analytical expressions for all wave functions and corresponding associated functions (non-diagonalizable Hamiltonian !) were obtained. Quite unexpectedly the same properties of the spectrum are present for two-dimensional anharmonic oscillator of some special form \cite{anharmonic} as well (see also \cite{bologna}).

It is known that in Hermitian Supersymmetric Quantum Mechanis with first order supercharges, the generalization \cite{abe} to higher dimensionality $d\geq 2$ leads \cite{kuru} to systems with separation of variables. But the situation is less restrictive and more interesting for the case of complex potentials - the first order operators $q$ may intertwine nontrivial Hamiltonians not amenable to separation of variables. Just this opportunity was realized in \cite{osc} and \cite{anharmonic} for two-dimensional models mentioned above.

In the present paper a class of three-dimensional models with oscillator-like shape invariance is studied. It is proved that the models are not amenable to conventional separation of variables for the complex potentials (Section 2), but their spectra are real and equidistant. The special case of such potential with quadratic interaction (Section 3) is solved completely. The symmetry properties of the model and degeneracy of energy levels are studied in Section 4. The Hamiltonian of the system is non-diagonalizable, and its wave functions and associated functions are built analytically in Sections 5 and 6.

\section*{\normalsize\bf 2.\quad Three-dimensional oscillator-like shape invariance: general solution.}

The intertwining relations for Hamiltonian with oscillator like shape invariance and with supercharges of first order in derivatives,
\ba
&&HA^+=A^+(H+2\lambda); \label{1}\\
&&A^+=a_i(\vec{x})\partial_i+a(\vec{x}); \quad H=-\partial_i^2+V(\vec{x});\quad \partial_i=\partial /\partial x_i;
\,\,\,i=1,2,3. \nonumber
\ea
are equivalent to the system of differential equations for the coefficient functions $a_i(\vec x),\, a(\vec x)$ and potential $V(\vec x):$
\ba
&&\partial_ka_i(\vec{x})+\partial_ia_k(\vec{x})=0; \label{2}\\
&&\Delta^{(3)}a_i(\vec{x})+2\partial_ia(\vec{x})=-2\lambda a_i(\vec{x});  \label{3}\\
&&\Delta^{(3)}a(\vec{x})+a_i(\vec{x})\partial_iV(\vec{x})=-2\lambda a(\vec{x}), \label{4}
\ea
where $\Delta^{(3)}$ is the three-dimensional Laplacian. It follows from (\ref{2}) that
\ba
\partial_1^2a_k(\vec{x})=\partial_2^2a_k(\vec{x})=\partial_3^2a_k(\vec{x})=0, \,\,\, k=1, 2, 3,\nonumber
\ea
and therefore, from (\ref{3}):
\ba
\partial_ia(\vec{x})=-\lambda a_i(\vec{x}). \nonumber
\ea
These equations are self-consistent if:
\ba
\partial_ia_k(\vec{x})=\partial_ka_i(\vec{x}), \,\,\, i\neq k. \nonumber
\ea
Thus, the solution of the system (\ref{2}) - (\ref{3}) is:
\ba
a_i=const; \quad a(\vec{x})=-\lambda a_ix_i,   \nonumber
\ea
where summation over repeated indices is implied, and possible constant in the r.h.s. can be eliminated by a suitable shift of coordinates. The remaining equation (\ref{4}) gives:
\ba
a_i\partial_i V(\vec{x})=2\lambda^2 a_i x_i.  \nonumber
\ea
Looking for potential $V(\vec x)$ in the form $V(\vec x) \equiv \lambda^2 \vec x^2 + v(\vec x),$ one has:
\ba
a_i\partial_iv(\vec{x})=0.  \label{10}
\ea
Obviously, at least two of the constants $a_i$ do not vanish, since otherwise variables in Hamiltonian are separated (systems with separation of variables are not the subject of this paper). For definiteness, let us take $a_1a_2\neq 0,$ leading to the general solution of (\ref{10}):
\ba
v(\vec{x})=u(a_2x_1-a_1x_2,a_3x_1-a_1x_3)     \nonumber
\ea
and the initial Hamiltonian of the form:
\ba
H=-\Delta^{(3)}+\lambda^2\vec{x}^2 + u(a_2x_1-a_1x_2,a_3x_1-a_1x_3).\label{12}
\ea

The form (\ref{12}) suggests the opportunity to separate variables in terms of
\ba
y_1=a_2x_1-a_1x_2, \quad y_2=a_3x_1-a_1x_3, \quad y_3=b_ix_i,  \label{13}
\ea
where $b_i$ are constants. Such transformation of variables is possible if the corresponding Jacobian
does not vanish:
\ba
D=b_1a_1^2+b_2a_1a_2+b_3a_1a_3\neq 0.  \nonumber
\ea
Rewriting the Hamiltonian in terms of $y_i:$
\ba
&&H(\vec{y})=-(a_1^2+a_2^2)\partial_{y_1}^2-(a_1^2+a_3^2)\partial_{y_2}^2-b_i^2\partial_{y_3}^2-
2a_2a_3\partial_{y_1}\partial_{y_2}-2(a_2b_1-a_1b_2)\partial_{y_1}\partial_{y_3}-\nonumber\\
&&-2(a_3b_1-a_1b_3)\partial_{y_2}\partial_{y_3}+\lambda^2x_k^2(\vec{y})+\Phi(y_1,y_2),   \label{15}
\ea
one obtains that separation is possible in the case of
\ba
b_2=\frac{a_2b_1}{a_1}, \,\,\, b_3=\frac{a_3b_1}{a_1}, \label{17}
\ea
so that $D=b_1a_i^2.$ Since if the constants $a_i$ are real, the Jacobian $D$ does not vanish, the interesting situation not amenable to separation of variables may appear only for $a_i^2=0,$ i.e. for complex values of $a_i$ ($b_1$ can not be chosen vanishing due to (\ref{17}), (\ref{13})). This is a reason why only the systems with complex potentials $(a_i^2=0)$ will be considered below (see the review papers \cite{bender-review}, \cite{mostafa-review}).

One can check that for $a_i, a(\vec x)$ defined above, the intertwining relation (\ref{1}) is fulfilled simultaneously with its partner relation, i.e.:
\ba
 [H, A^{\pm}]=\pm 2\lambda A^{\pm},  \label{18}
\ea
where $A^{\pm}$ are defined as:
\ba
 A^{\pm}=a_i\partial_i\mp\lambda a_ix_i    \label{A}
\ea
(one has to notice that the operators $A^{\pm}$ are not necessarily mutually Hermitian conjugated).
It is clear from (\ref{18}) that $A^{\pm}$ increase and decrease the energy by $2\lambda ,$ correspondingly. If the ground state of the model (with zero energy) exists, it is defined by two equations:
\ba
&&A^-\Psi_0=(a_i\partial_i+\lambda a_ix_i)\Psi_0=0,\label{19}\\
&&H\Psi_0=E_0\Psi_0;\quad E_0\equiv 0.\label{20}
\ea

It is convenient to look for the solution of (\ref{19}) in the form:
\ba
\Psi_0\equiv\exp(-\lambda\vec{x}^2/2)\psi_0(\vec{x}).\label{21}
\ea
After substitution into (\ref{19}), the new function $\psi_0$ satisfies
$$a_i\partial_i\psi_0(\vec{x})=0,$$
and its solution depends on two combinations of coordinates:
$$\psi_0(\vec{x})=\psi_0(a_2x_1-a_1x_2,a_3x_1-a_1x_3); \quad \Psi_0=\exp(-\lambda\vec{x}^2/2)\psi_0(a_2x_1-a_1x_2,a_3x_1-a_1x_3).$$
According to (\ref{20}), $\psi_0(a_2x_1-a_1x_2,a_3x_1-a_1x_3)$ has to satisfy:
\ba
-\Delta^{(3)}\psi_0+2\lambda x_i\partial_i\psi_0 + u\psi_0 = -3\lambda\psi_0,\label{22}
\ea
which is difficult to solve for $\psi_0$ with an arbitrary function $u.$ We will act vice versa, expressing $u$ in terms of given $\psi_0 .$ For this purpose, let us write $\psi_0 $ in the exponential form:
$$\psi_0\equiv\exp W(y_1,y_2),$$
and then from (\ref{22}), we find that:
\ba
u=(\partial_i W(\vec{x}))^2+\Delta^{(3)}W(\vec{x})-2\lambda x_i\partial_i W(\vec x)-3\lambda .\label{23}
\ea

Thus, the Hamiltonian
\ba
H=-\Delta^{(3)}+\lambda^2\vec{x}^2+(\partial_i W(\vec{x}))^2+\Delta^{(3)}W(\vec{x})-
2\lambda x_i(\partial_i W)-3\lambda \label{25}
\ea
with arbitrary function $W$ has the ground state:
\ba
\Psi_0=\exp(-\lambda\vec{x}^2/2+W(a_2x_1-a_1x_2,a_3x_1-a_1x_3))\label{26}
\ea
with zero energy. The excited states with energies $E_n=2\lambda n$ for the considered case $a_i^2=0$ can be built by the "creation" operator $A^+$:
\ba
\Psi_n=(A^+)^n\Psi_0=(-2\lambda)^n(a_ix_i)^n\Psi_0.\label{27}
\ea
From the definition (\ref{A}) one has:
\ba
[A^+,A^-]=2\lambda a_i^2=0,  \label{28}
\ea
and therefore, not only the ground state wave function but also all excited wave functions are annihilated by $A^-:$
\ba
A^-\Psi_n=0.\label{30}
\ea
The full analysis of the general model (\ref{25}) seems to be too difficult a task. In particular, there is no guarantee that the states (\ref{27}) exhaust the entire set of excited states. For this reason, in the next Section a particular case of the model will be studied in more details.

\section*{\normalsize\bf 3.\quad The particular case.}

Let us consider here the particular case of parameters of the model:  $a_3=0, a_1=1, a_2=-i,$ and the simple form of
function $W$ which leads to the quadratic interaction:
$$W(a_2x_1-a_1x_2,x_3)=g(x_1-ix_2)x_3=g\bar{z}x_3,$$
where $z\equiv x_1+ix_2\, \bar z\equiv x_1-ix_2.$
Then, from (\ref{A}), (\ref{25}), (\ref{26}) one has:
\ba
&&H=-\partial_i^2+\lambda^2x_i^2+g^2\bar{z}^2-4\lambda g\bar{z}x_3-3\lambda,\label{31}\\
&&A^{\pm}=2\partial_z\mp\lambda\bar{z},\label{AA}\\
&&\Psi_0=\exp(-\lambda\vec{x}^2/2+g\bar{z}x_3), \label{F}
\ea
where the mutual independence of $z,\, \bar z$ provides (\ref{28}), and for $\lambda > |g|$ the wave function is exponentially decreasing at infinity.

Actually, this system reproduces exactly one of the three-dimensional models of a recent paper \cite{three}, if
in Eq.(35) of that paper $b\equiv 2\lambda, b_3\equiv 2g,\, x_2\to -x_2$ and $ Q^{\pm} \leftrightarrow Q^{\mp}.$
Thus the same Hamiltonian (\ref{31}) is intertwined by supercharges $Q^{\pm}$ of second order in derivatives:
\ba
[H,Q^{\pm}]=\pm 4\lambda Q^{\pm},\label{32}
\ea
where
\ba
&&Q^-=\partial_1^2+\partial_2^2-\partial_3^2+C_i\partial_i+B,\nonumber\\
&&Q^+=\partial_1^2+\partial_2^2-\partial_3^2-C_i\partial_i+B-\partial_iC_i,\nonumber\\
&&C_1=2(\lambda x_1-gx_3),\nonumber\\
&&C_2=-2(\lambda x_2+igx_3),
\nonumber\\
&&C_3=2(-\lambda x_3+g\bar{z}),\nonumber\\
&&B=\lambda^2(z\bar{z}-x_3^2)-g^2\bar{z}^2+\lambda,\nonumber\\
&&C_i\partial_i=2\lambda (z\partial_z+\bar{z}\partial_{\bar{z}}-x_3\partial_3)-
4gx_3\partial_z+2g\bar{z}\partial_3.\nonumber
\ea

The intertwining relations (\ref{32}) also have the form of oscillator-like shape invariance of Hamiltonian $H,$
but with the double spacing between energy levels. The minimal energy state corresponds to the zero mode of $Q^-,$
and it can be easily found. Indeed, direct calculation shows that:
\ba
&&Q^-=\Psi_0(\partial_1^2+\partial_2^2-\partial_3^2)\Psi_0^{-1},\nonumber\\
&&Q^+=Q^--2\lambda -2C_i\partial_i, \nonumber
\ea
therefore, the same ground state $\Psi_0$ given in (\ref{F}) is annihilated not only by $A^-,$ but also by $Q^-.$ One can check that other possible zero modes of $Q^-$ can not be simultaneously the wave functions of $H.$

It is clear that both $(A^+)^2\Psi_0$ and $Q^+\Psi_0$ are the wave functions of the Hamiltonian $H$ with the same energy $4\lambda .$ The calculation shows that
\ba
Q^+\Psi_0=-2(\lambda\Psi_0+C_i\partial_i\Psi_0)=-2\lambda[1+2\lambda(x_3^2-z\bar{z})]\Psi_0-
8\lambda g^2\bar{z}^2\Psi_0,\nonumber
\ea
i.e. there is degeneracy at the second excited level $4\lambda$ with two independent wave functions:
\ba
\Psi_2=(A^+)^2\Psi_0\sim \bar{z}^2\Psi_0,\,\,\,
\widetilde{\Psi}_2=[1+2\lambda(x_3^2-z\bar{z})]\Psi_0. \nonumber
\ea
Analogously, higher wave functions of the form $Q^+\Psi_n(\vec x)$ can be calculated, using expression (\ref{27}),
which, in our case of $a_1=1,\,\, a_2=-i,\,a_3=0 ,$ means $\Psi_n \sim (\bar z)^n\Psi_0:$
\ba
&&Q^+\bar{z}^n\Psi_0=(Q^+-2\lambda -2C_i\partial_i)\bar{z}^n\Psi_0=-2\lambda\bar{z}^n\Psi_0
-2\bar{z}^nC_i\partial_i\Psi_0-2\Psi_0C_i\partial_i\bar{z}^n=\nonumber\\
&&=-2\lambda[2n+1+2\lambda(x_3^2-z\bar{z})]\bar{z}^n\Psi_0-8\lambda g^2\bar{z}^{2+n}\Psi_0.\nonumber
\ea
Therefore, for arbitrary $n,$ the function
\ba
\widetilde{\Psi}_n\equiv [2n+1+2\lambda(x_3^2-z\bar{z})]\bar{z}^n\Psi_0 \nonumber
\ea
is the wave function of $H$ with energy value $2\lambda n+4\lambda,$ being degenerated with $\Psi_{n+2}.$
This is the general situation: action of operators $Q^+$ leads to degeneracy of levels. Degenerate wave functions are:
\ba
\Psi_{kn}(\vec x)=(Q^+)^k(A^+)^n\Psi_0;\quad \Psi_{0n}\equiv \Psi_n; \,\,\, E_{kn}=2\lambda(n+2k),\label{56}
\ea
with the multiplicity of level $N=k+1+[n/2].$ It is necessary to mention that the order of "creation" operators in (\ref{56}) is irrelevant due to following commutation relations between operators $Q^{\pm}$ and $A^{\pm}:$
\ba
&&[A^+, Q^-]=4\lambda A^-, \,\,\, [A^-, Q^+]=-4\lambda A^+;\label{39}\\
&&[A^+, Q^+]=[A^-, Q^-]=0.\label{40}
\ea

One may expect that combinations $A^{\pm}$ and $Q^{\pm}$ of third order in derivatives will give additional degeneracy of some levels, since they also satisfy oscillator-like shape invariant intertwining relations obtained from (\ref{18}) and (\ref{32}):
\ba
&&[H, A^+Q^-]=-2\lambda A^+Q^-, \,\,\, [H, Q^-A^+]=-2\lambda Q^-A^+; \label{41}\\
&&[H, A^-Q^+]=2\lambda A^-Q^+, \,\,\, [H, Q^+A^-]=2\lambda Q^+A^-; \label{42}\\
&&[H, A^+Q^+]=6\lambda A^+Q^+, \,\,\, [H, A^-Q^-]=-6\lambda A^-Q^-. \label{43}
\ea
However, additional degeneracy is not generated, since:
\ba
&&A^+Q^-\Psi_n=A^+\Psi_0(\partial_1^2+\partial_2^2-\partial_3^2)\bar{z}^n=0,\nonumber\\
&&Q^-A^+\Psi_n=(A^+Q^--4\lambda A^-)\Psi_n=0, \nonumber\\
&&A^-Q^+\Psi_n=A^-(Q^--2\lambda -2C_i\partial_i)\Psi_n=\nonumber\\
&&=-2A^-C_i\partial_i\Psi_n=-2(C_i\partial_i A^-+2\lambda A^+)\Psi_n=
-4\lambda A^+\Psi_n, \nonumber\\
&&Q^+A^-\Psi_n=0, \nonumber\\
\ea
i.e. action of these third order operators does not give any new states.

\section*{\normalsize\bf 4.\quad Symmetries.}

The degeneracy of levels which was described in the previous Section indicates the existence of some symmetry of Hamiltonian. Indeed, the commutation relations (\ref{18}), (\ref{32}) lead immediately to the symmetry operators commuting with the Hamiltonian (notations will be clear below):
\ba
R_0=A^+A^-=A^-A^+;\quad \widetilde R_1=\frac{1}{2}[Q^+, Q^-]; \quad \widetilde R_2=Q^+Q^-.
\label{R}
\ea
These operators are bilinear in "creation" and "annihilation" operators, and they definitely do not change the energy of the state. The first of these operators is of second order in derivatives, and the others two are of fourth order, but the order of $\widetilde R_1$ can be reduced by two units, since:
\ba
\widetilde{R}_1=-4\lambda H+16g\partial_z\partial_3+8g\lambda \bar{z}(g\bar{z}-\lambda x_3) -12\lambda^2.\nonumber
\ea
Thus, a more convenient symmetry operator instead of $\widetilde R_1$ is
\ba
R_1=2\partial_z\partial_3+\lambda \bar{z}(g\bar{z}-\lambda x_3).\label{38}
\ea
Also, the fourth order symmetry $\widetilde R_2$ can be replaced by third order symmetry operator which is proportional to the commutator $[R_0, \widetilde R_2].$ Relations (\ref{39}), (\ref{40}) help to express this operator as:
\ba
R_2=4\lambda\bar{z}\partial_z Q^--(\lambda+C_i\partial_i)(A^-)^2.\label{R2}
\ea
One can calculate the mutual commutators between the symmetry operators $R_0, R_1, R_2.$ They are:
\ba
[R_0, R_1]=0;\quad [R_2, R_0] = 4\lambda R_0^2;\quad [R_2, R_1] = -2gR_0^2. \label{commut}
\ea

More complicated symmetry operators consist of the products of $A^{\pm}$ and $Q^{\pm}$ such that the energy of state is not changed:
\be
R_3=Q^+(A^-)^2;\quad R_4=Q^-(A^+)^2.
\label{r1}
\ee
Using the explicit expressions for $A^{\pm},\, Q^{\pm}$ and commutation relations (\ref{39}), (\ref{40}), one can find that:
\be
R_3-R_4=2R_2+8\lambda R_0,\nonumber
\ee
so that only one of them - for example, $R_3$ - can be taken as independent. Its nonvanishing commutators with $R_0,\, R_1,\, R_2$
do not give any new independent symmetry operator. The straightforward calculations prove that higher symmetry operators of the form $(A^+)^{2n}(Q^-)^n,\, (A^-)^{2n}(Q^+)^n$ as well as the products with different order of $A$ and $Q$ also do not provide new independent symmetries.

In the context of conventional - Hermitian - Quantum Mechanics, such situation would be called as maximally superintegrable \cite{winternitz}. Indeed, for the three-dimensional system one has two mutually commuting symmetry operators $R_0,\, R_1$
(complete integrability) and, in addition, two symmetry operators $R_2,\, R_3,$ which do not commute with them. Commutators of this last operator with polynomials of $R_0,\,R_1$ are not functionally independent from $R_0$ and do not provide additional symmetries.

Now, it is instructive to show the connection of symmetries with degeneracy of energy levels. Using commutation relations between operators $A^{\pm}, \, Q^{\pm}$ and definitions of $R_0,\, R_1,\, R_2,$ one finds the action of the symmetry operators onto wave functions $\Psi_{k n},$ (\ref{56}):
\ba
R_0\Psi_{kn}&=&-4\lambda k\Psi_{(k-1)(n+2)};\label{59}\\
R_1\Psi_{kn}&=&2gk\Psi_{(k-1)(n+2)};\label{60}\\
R_2\Psi_{kn}&=&8g^2k(k-1)\Psi_{(k-2)(n+4)}-4\lambda^2k(2n+3)\Psi_{(k-1)(n+2)};\label{66}\\
R_3\Psi_{kn}&=&16\lambda^2k(k-1)\Psi_{(k-1)(n+2)}.\label{666}
\ea
It is evident that all wave functions in (\ref{59}) - (\ref{666}) correspond to the same level with energy $E_{kn}=2\lambda (n+2k).$ The repeated actions of symmetry operators span all wave functions which belong to this degenerate level.

\section*{\normalsize\bf 5.\quad Norms, scalar products and non-diagonalizability of Hamiltonian.}

The Hamiltonian (\ref{31}) is not Hermitian, but it obeys the property of pseudo-Hermiticity \cite{bender}:
\ba
\eta H \eta^{-1} = H^{\dagger}
\label{pseudo}
\ea
with $\eta $ an invertible Hermitian operator. In the present case, an operator $\eta $ can be chosen as $\eta = P_2,$ the inversion $x_2 \to - x_2.$ The theory of one-dimensional pseudo-Hermitian systems was developed during last decade \cite{bender}, the pseudo-Hermitian models in two dimensions were studied in \cite{pseudo}, \cite{osc}, \cite{anharmonic}, where the basic formulas can be found. In particular, Quantum Mechanics for such systems with unbroken $\eta T=P_2T -$symmetry $P_2T\Psi_n(\vec x)=\Psi_n(\vec x)$ is built with the new scalar product ($\eta -$product), which in the present case has the form:
\ba
<\Psi \mid\eta\mid \Phi> = <<\Psi \mid \Phi>> = \int\Psi\Phi,
\label{scalar}
\ea
instead of the standard $\int\Psi^*\Phi.$

With the scalar product (\ref{scalar}), the norm of ground state $\Psi_0$ is:
\ba
&&\int\Psi_0^2d^3x=\int\exp(-\lambda x_i^2+2g\bar{z}x_3)d^3x=
\frac{1}{2}\int\exp(-\lambda z\bar{z})dzd\bar{z}\int\exp(-\lambda x_3^2+2g\bar{z}x_3)dx_3=\nonumber\\
&&=\frac{1}{2}\sqrt{\frac{\pi}{\lambda}}\int\exp(-\lambda z\bar{z}+
\frac{g^2\bar{z}^2}{\lambda})dzd\bar{z}=\sqrt{(\pi/\lambda)^3}.\label{71}
\ea
The norms of excited states $\Psi_n$ from (\ref{27}) can be also calculated:
\ba
&&<<\Psi_n \mid \Psi_n>>=<<(A^+)^n\Psi_0 \mid (A^+)^n\Psi_0>>=(-1)^n<<\Psi_0 \mid R_0^n\Psi_0>>=\nonumber\\
&&=(-1)^n<<\Psi_0 \mid \Psi_0>>\delta_{n0}.\label{n}
\ea

In turn, the norms of excited states $\Psi_{kn}=(Q^+)^k\Psi_n$ given by (\ref{56}) can be found as well, although with more involved calculations using $R_1\Psi_n=0;\, Q^-\Psi_n=0$ and the commutation relations between $Q^{\pm}:$
\ba
[Q^+, Q^-]=-8\lambda (H+3\lambda)+16gR_1.\label{63}
\ea
One can check that required scalar products are:
\ba
<<\Psi_{kn} \mid \Psi_{kn}>> = <<(Q^+)^k\Psi_n \mid \Psi_{kn}>> = <<\Psi_n \mid (Q^-)^k\Psi_{kn}>> \sim <<\Psi_n \mid \Psi_n>>\sim\delta_{n0},   \label{73}
\ea
where the coefficients in the last step are not important. Analogously, one can check that for different values of indices $k,\, n,$ the scalar products
\ba
<<\Psi_{kn} \mid \Psi_{k'n'}>> \sim \delta_{kk'}\delta_{nn'}\delta_{n0}.   \label{733}
\ea

Thus, all wave functions $\Psi_{kn}$ of the model, excluding the ground state $\Psi_0$ and the states $\Psi_{k0},$ have zero norms and are orthogonal to each other. The self-orthogonality of wave functions signals that we deal with a non-diagonalizable Hamiltonians. This situation was studied recently in one-dimensional \cite{non-diag} and two-dimensional \cite{osc}, \cite{anharmonic} context, where necessary technical details can be found. Non-diagonalizability means that each wave function $\Psi_{kn}, \, n>0$ with zero norm must be accompanied with a set of associated functions which participate in the resolution of identity.

\section*{\normalsize\bf 6.\quad Associated functions.}

Referring back to the papers \cite{non-diag}, \cite{osc}, \cite{anharmonic} for details on the structure of non-diagonalizable non-Hermitian Hamiltonians, biorthogonal basis, scalar products between wave functions and associated functions and nontrivial form of resolution of unity, let us remind the most important relations which must be fulfilled.

For such systems, each self-orthogonal wave function $\Psi_{kn}(\vec x)$ with zero norm (below they be denoted as $\Psi_{kn,0}(\vec x)$) must be accompanied with a set of $p_{kn}-1$ associated functions $\Psi_{kn,m},\, m=1,2,...,p_{kn}-1,$ where $p_{kn}$ is called the dimension of Jordan cell. This situation should be distinguished from the conventional degeneracy of the energy level (here this degeneracy also exists). A new notation with an additional index of wave function $\Psi_{kn}\equiv\Psi_{kn,0}$ is useful, since by definition these functions obey:
\begin{equation}\label{assoc}
 (H-E_{kn})\Psi_{kn,m}=\Psi_{kn,m-1},\quad m=1,2,...,p_{kn}-1;\quad (H-E_{kn})\Psi_{kn,0}=0,
\end{equation}
where all functions are supposed to be normalizable according to the scalar product (\ref{scalar}).
Thus, each self-orthogonal eigenfunction $\Psi_{kn,0}$
is supposed to be accompanied by a set of associated functions $\Psi_{kn,m},\, m=1,2,...,p_{kn}-1.$

With these notations, according to the general formalism which was given in detail for some one-dimensional
models \cite{non-diag} and two-dimensional models \cite{osc}, \cite{anharmonic}, the scalar products must be:
\ba
\langle\langle\Psi_{kn,m}|\Psi_{k'n',m'}\rangle\rangle =
\int \Psi_{kn,m}(\vec x)\Psi_{k'n',m'}(\vec x)d^3x =\delta_{kk'}\delta_{nn'}\delta_{m\, (p_{kn}-m'-1)};\,\, m'=0,1,2,...,p_{kn}-1\label{product}
\ea

The Hamiltonian $H$ is clearly non-diagonal, but block-diagonal. Each block - Jordan cell - has dimensionality $p_{kn},$ which will be shown (see Eq.(\ref{final-final})) to equal:
\ba
p_{kn}=n+1.
\label{pkn}
\ea
The self-orthogonality of wave functions - $\Psi_{kn,0},\, n\neq 0$ - was already demonstrated in (\ref{73}), (\ref{n}). To complete the construction of the Jordan cell, it is necessary to find the corresponding associated functions with properties (\ref{assoc}), (\ref{product}) above. The procedure is rather similar to that in \cite{osc}, \cite{anharmonic}.

We will start from $\Psi_{kn,0}$ with $k=0.$ Due to relations
\begin{equation}
(A^-)^k(H-E_n)=(H-E_{n-k})(A^-)^k,   \nonumber
\end{equation}
the associated functions $\Psi_{0n,m}$ for corresponding wave functions
$$\Psi_{0n,0}=(A^+)^n\Psi_{00,0}=c_{n,0}\bar z^n\exp{(-\frac{\lambda x_i^2}{2}+g\bar z x_3)}$$
in (\ref{56}) ($c_{n,0}$ are constants) satisfy, as in \cite{osc}, \cite{anharmonic}, the equations:
\ba
(A^-)^m\Psi_{0n,m}=2^m a_{n,m}\Psi_{0(n-m),0},\nonumber
\ea
where $a_{n,m}$ constants. The solution is searched in the form:
\ba
\Psi_{0 n,m}=\exp(-\frac{\lambda x_i^2}{2}+g\bar{z}x_3)\varphi_{n,m}, \nonumber
\ea
where the function $\varphi_{n,m}$ has to satisfy:
\ba
\partial_z^m\varphi_{n,m}=a_{n,m}c_{n-m,0}\bar{z}^{n-m}. \nonumber
\ea
Its general solution is:
\ba
\varphi_{n,m}=\frac{a_{n,m}c_{n-m,0}}{m!}z^m\bar{z}^{n-m}+
\sum_{i=0}^{m-1}g_{n,m}^{(i)}(\bar{z},x_3)z^i,\nonumber
\ea
where functions $g_{n,m}^{(i)}(\bar{z},x_3)$ can be found from equations:
\ba
(H-E_n)\Psi_{0 n,m}=\Psi_{0 n,m-1}\label{a6}
\ea
by direct recurrence procedure. An alternative way is to calculate the associated functions for the lowest values of $n,$ and to guess their form for higher values. Indeed, one can prove by induction that
\ba
\Psi_{0 n,m}\sim (A^+)^{n-m}\exp(-\frac{\lambda x_i^2}{2}+g\bar{z}x_3)(z-\frac{2g}{\lambda}x_3)^m
\sim (A^+)^{n-m}\partial^m_{\bar{z}}\Psi_{0,0}, m=0, 1, ..., n   \nonumber
\ea
satisfies (\ref{assoc}). This expression leads (by straightforward calculations) to the necessary scalar products:
\ba
(\Psi_{0 n,m},\Psi_{0 n,m'})\sim \delta_{m (n-m')}.   \nonumber
\ea

In turn, the associated functions $\Psi_{k n,m}$ for $k \neq 0$ satisfy:
\ba
(H-E_{kn})\Psi_{kn,m}=\Psi_{kn,m-1},\,\,E_{kn}=2\lambda (n+2k).\label{a39}
\ea
Taking into account that
\ba
(Q^+)^k(H-E_{0n})=(H-E_{kn})(Q^+)^k,\label{a40}
\ea
one can try to identify:
\ba
\Psi_{kn,m}=(Q^+)^k\Psi_{0 n,m}.   \nonumber
\ea
It is easy to check that
\ba
&&<<\Psi_{kn,m}\mid\Psi_{kn,0}>> = <<(Q^+)^k\Psi_{0 n,m} \mid \Psi_{kn,0}>> =\nonumber\\&&= <<\Psi_{0 n,m} \mid (Q^-)^k\Psi_{kn,0}>> \sim <<\Psi_{0 n,m} \mid \Psi_{0 n,0}>> \sim \delta_{nm}.\label{a43}
\ea
i. e. $\Psi_{kn,0}$ is orthogonal to all $\Psi_{kn,m}$ with $ m\neq n.$
In order to satisfy the necessary conditions for all scalar products, one has to use the freedom to add the solution
$\Psi_{kn, 0}$ of the homogeneous equation (\ref{assoc}) with arbitrary constant coefficient. This is enough to fulfill the conditions:
\ba
<<\Psi_{kn,m}\mid \Psi_{kn,m'}>> \sim \delta_{m (n-m')},   \nonumber
\ea
which mean that we deal with Jordan cells of matrix dimensionalities $p_{kn}=n+1.$
Analogous calculations show that:
\ba
<<\Psi_{kn,m}\mid \Psi_{k'n',m'}>> =\delta_{nn'}\delta_{kk'}\delta_{m (n-m')},
\label{final-final}
\ea
completing the proof that the Hamiltonian is block-diagonal with Jordan cells of dimensionality $p_{kn}=n+1.$

\section*{\normalsize\bf 7.\quad Conclusions.}

In the present paper the method of supersymmetric shape invariance was applied to investigate a new model. This model has several interesting properties which distinguish it from others recently studied \cite{osc}, \cite{anharmonic}. First of all, it is three-dimensional, and although the interaction is quadratic, the special choice of coupling constants with one of them pure imaginary does not allow to separate variables. Similarly to two-dimensional models of \cite{osc}, \cite{anharmonic}, the Hamiltonian is non-diagonalizable, but now it incorporates nontrivial symmetries. Two of the symmetry operators are mutually commuting, while two others are not. Besides the block-diagonal form of Hamiltonian with Jordan cells corresponding to wave functions, each energy level is degenerated due to symmetry operators which mix the wave functions with the same energy. The number of symmetry operators just implies that our model belongs to a class of maximally superintegrable models in conventional Hermitian Quantum Mechanics, but we do not know of general statements on possible numbers of symmetries in non-Hermitian case. The analysis above may lead, in particular, to some hints concerning such general theorems.

\section*{\bf Acknowledgments.}

The work of M.V.I. and D.N.N. was partially supported by INFN and the University of Bologna.

\end{document}